# Generalizability of Deep Adult Lung Segmentation Models to the Pediatric Population: A Retrospective Study

**Sivaramakrishnan Rajaraman[1,\*], Feng Yang[1], Ghada Zamzmi[1], Zhiyun Xue[1], Sameer Antani[1]**

[1] Computational Health Research Branch, National Library of Medicine, National Institutes of Health, Bethesda, MD 20894, USA

## ABSTRACT

Lung segmentation in chest X-rays (CXRs) is an important prerequisite for improving the specificity of diagnoses of cardiopulmonary diseases in a clinical decision support system. Current deep learning models for lung segmentation are trained and evaluated on CXR datasets in which the radiographic projections are captured predominantly from the adult population. However, the shape of the lungs is reported to be significantly different across the developmental stages from infancy to adulthood. This might result in age-related data domain shifts that would adversely impact lung segmentation performance when the models trained on the adult population are deployed for pediatric lung segmentation. In this work, our goal is to (i) analyze the generalizability of deep adult lung segmentation models to the pediatric population and (ii) improve performance through a stage-wise, systematic approach consisting of CXR modality-specific weight initializations, stacked ensembles, and an ensemble of stacked ensembles. To evaluate segmentation performance and generalizability, novel evaluation metrics consisting of mean lung contour distance (MLCD) and average hash score (AHS) are proposed in addition to the multi-scale structural similarity index measure (MS-SSIM), the intersection of union (IoU), Dice score, 95% Hausdorff distance (HD95), and average symmetric surface distance (ASSD). Our results showed a significant improvement ($p < 0.05$) in cross-domain generalization through our approach. This study could serve as a paradigm to analyze the cross-domain generalizability of deep segmentation models for other medical imaging modalities and applications.

# 1. INTRODUCTION

Lung segmentation is an important pre-requisite step for deep learning (DL)-based clinical decision support system to facilitate early diagnosis of cardio-pulmonary diseases including tuberculosis (TB), pneumonia, and lung cancer, among others (Osadebey et al., 2021). DL models deliver superior performance in medical image semantic segmentation tasks, particularly for segmenting the lung regions of interest (ROIs) in chest X-rays (CXRs) (Candemir & Antani, 2019; Liu et al., 2022). These studies report segmentation performance using models that were trained and evaluated on the widely used Shenzhen (Jaeger et al., 2014), Montgomery (Jaeger et al., 2014), and Japanese Society of Radiological Technology (JSRT) (Oda et al., 2009) CXR datasets. However, the images in these three datasets are primarily of adults. Unlike adults, infants have smaller, triangular-shaped lungs (Arthur, 2000). Their lungs show relatively shallower costophrenic angles with elevated diaphragms, and the mediastinum and cardiac contour tend to be relatively larger compared to adults. These differences in lung shapes across the pediatric developmental stages are qualitatively and quantitatively evaluated in the literature (Candemir et al., 2015), where the authors devised three age groups, viz., group 1 (1 day to < 24 months), group 2 (24 months to < 11 years), and group 3 (11 years to < 18 years), respectively, based on the changes in the lung shape from infancy to adulthood, and emphasized the need for age-sensitive lung segmentation models.

Generalizability in DL refers to the ability of a trained DL model to accurately make predictions on new, unseen data. In other words, a model is said to be generalizable if it can effectively learn the underlying patterns and relationships in a dataset and apply that knowledge to make accurate predictions on previously unseen data. Domain generalization is an important aspect of DL, and it is particularly useful in the context of medical image analysis. Domain generalization aims to integrate the knowledge learned from multiple source domains to reduce deviations in acquired knowledge due to inter-domain differences. It can help improve the accuracy and reliability of medical image analysis, which is critical for the diagnosis and treatment of diseases, and it increases the practical utility of DL models. Age-related domain generalization refers to the ability of the DL models to generalize across different age groups. Specifically, it refers to the ability of these models to maintain good performance on tasks even when the training and test data come from different age groups.

A study of the literature reveals that the majority of DL models trained on internal/source domains may not generalize to the unseen external/target domains due to differences in the distribution (Xin et al., 2022) of individual data elements. Medical image data like CXRs can be highly variable, with differences in image acquisition protocols, patient heterogeneity based on age, gender, race, and ethnicity, and varying characteristics of the underlying disease manifestations (Suzuki, 2017). These variations can cause significant challenges when building DL models that can generalize to new data from different domains. Resulting domain shift issues lead to performance degradation and sub-optimal generalization. Further, this domain shift might lead to poor performance, if not failure, when these models are deployed in expert systems, which can potentially degrade people's trust in the model, and threaten clinical decisions and best-practice recommendations.

This study aims to analyze age-related domain generalization, i.e., the generalizability of deep adult lung segmentation models to the pediatric population. Fig. 1 illustrates the workflow of age-related domain generalization in the context of our study. Here, the widely used Shenzhen, Montgomery, and JSRT CXR collections represent the multiple source domains and a private pediatric CXR collection (Candemir et al., 2015) represents the unseen target domain. We aim to evaluate age-related domain generalization, i.e., analyzing if the DL models trained on multiple source domains containing the CXRs captured from the adult population would generalize to an unseen pediatric CXR collection.

## 2. Related Literature and contributions of the study

Recent research has demonstrated the importance of domain generalization in medical image analysis and provided insight into the techniques used to achieve domain generalization in DL models, particularly when trained and evaluated on medical imaging datasets that often have limited sample sizes, high levels of variability, and complex underlying structures. A novel transfer learning method based on instance-level adaptation was proposed in the literature (Zheng et al., 2021) to improve generalization. With limited data availability, the proposal demonstrated superior performance on unseen target domains compared to conventional transfer learning methods. In another study (Lyu et al., 2022), the authors proposed a method to manipulate the data and improve domain generalization. The data was augmented to generate novel imaging domains that improved the training set's diversity. In another study (Stacke et al., 2021), the authors measured the representation shift that quantified the magnitude of domain shift in a tumor classification task. The authors observed that the representation shift correlated with reduced performance during cross-domain evaluation and helped evaluate the sensitivity of the model to domain variations.

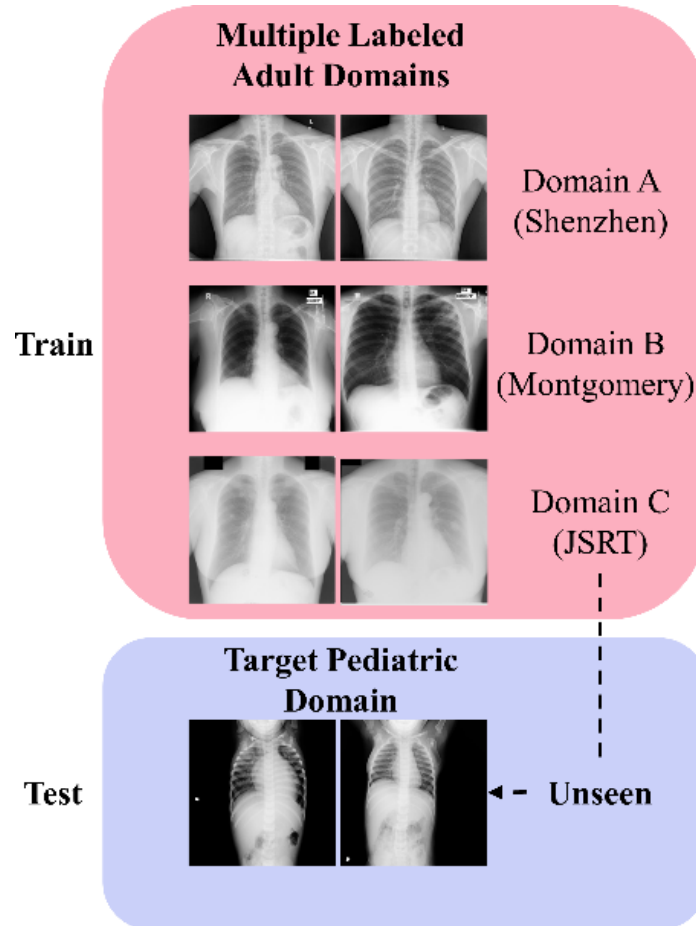

Fig. 1 – Illustrating the workflow for age-related domain generalization in the context of our study.

A variety of domain generalization solutions are proposed in the literature in the context of CXR image analysis. In (Cohen et al., 2020), the authors empirically demonstrated model generalization issues due to label shifts. They observed that the external test data was inadequately represented in the training and even high-performing models disagreed with their predictions. In another study (Zech et al., 2018), the authors trained CNN-based models on publicly available CXR datasets to classify them as showing pneumonia-consistent manifestations or normal lungs. A significant reduction in performance was reported using cross-institutional test sets compared to the internal test sets. In a recent study (Xin et al., 2022), the authors trained DL models to classify the pediatric CXRs as showing pneumonia-consistent manifestations or normal lungs and evaluated model generalization using external test sets. They observed that the models delivered a superior performance with the internal test set but had a significantly lower performance with an external test set. A semi-supervised regularization method was proposed in the literature (Zhang et al., 2022). This method used the stability and orthogonality of the learned features to alleviate domain generalization issues in a CXR classification task. All these prior works demonstrate the importance of domain generalization in medical imaging and propose innovative methods for improving the generalization performance of DL models across different domains.

To the best of our knowledge, this proposal is the first to analyze and improve age-related domain generalization for a CXR-based segmentation task, i.e., improve lung segmentation performance from models initially trained on adults by generalizing them to include pediatric images. The DL models are evaluated on the unseen pediatric test through a stage-wise systematic approach as illustrated in Fig. 2 and discussed below:

(i)     A private collection of pediatric CXRs (Candemir et al., 2015) was split into three age groups (P1, P2, and P3) as discussed in the literature (Candemir et al., 2015).

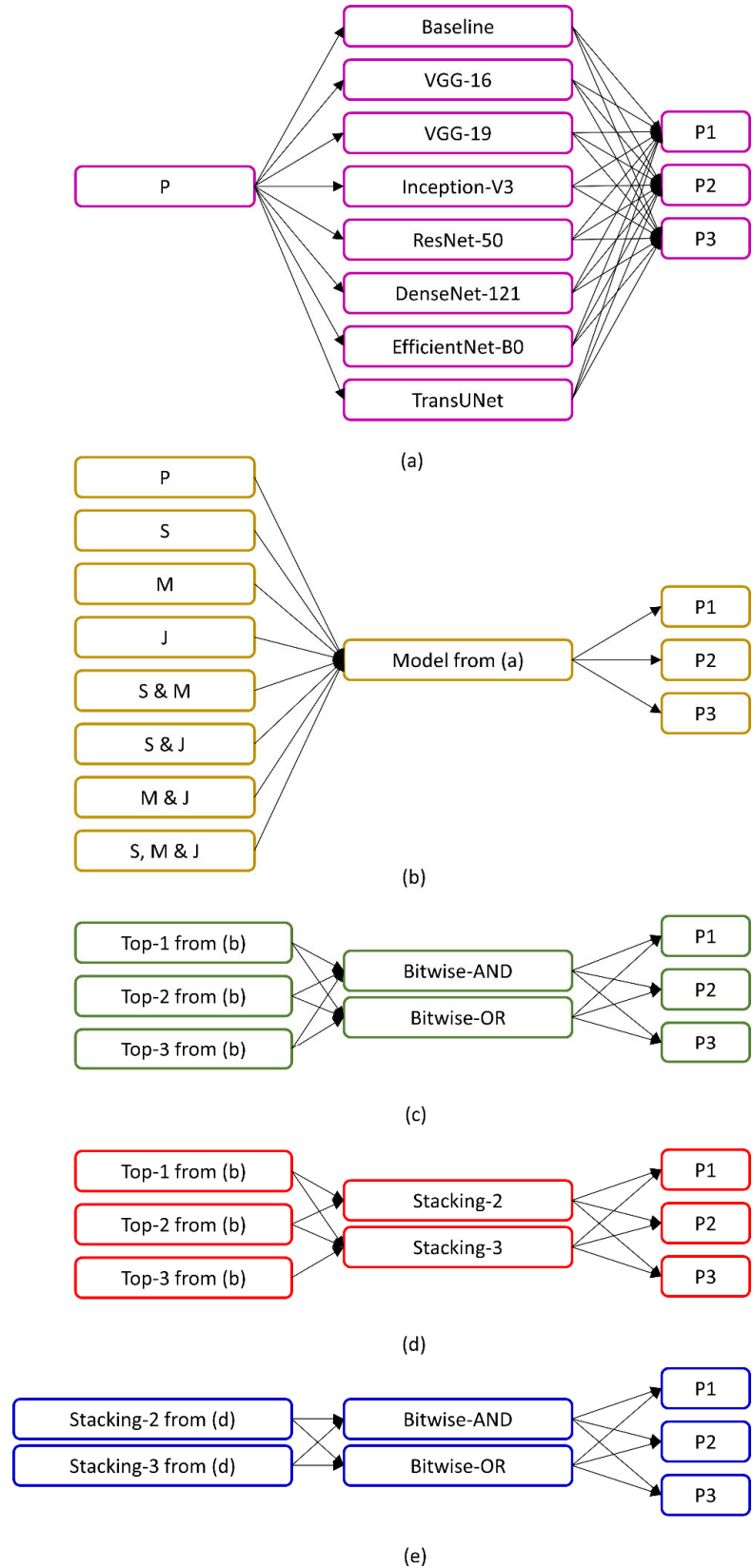

(a)

(b)

(c)

(d)

(e)

Fig. 2 – Various steps involved in the proposed stage-wise systematic approach. The baseline denotes the conventional UNet architecture (Ronneberger et al., 2015). The acronyms P, S, M, and J denote the Pediatric-N, Shenzhen-N, Montgomery-N, and JSRT-N CXR datasets. (a) The UNet and TransUNet models were trained on the P CXR dataset and tested with the three pediatric

test groups P1, P2, and P3. (b) The best-performing model architecture from (a) was trained on S, M, and J adult CXR datasets in isolation and their combinations and tested with P1, P2, and P3. (c) and (d) The predictions of the top-K (K = 2, 3) adult models from (b) were aggregated through bitwise operations and a stacked ensemble, respectively. (e) An ensemble of stacked ensembles was constructed where the predictions of the stacked ensembles constructed using the top-2 and top-3 adult models from (d) were aggregated using bitwise operations.

(i)     Segmentation models consisting of the conventional UNet, (Ronneberger et al., 2015), its variants with several ImageNet pretrained encoder backbones (Iakubovskii, 2019), and the conventional TransUNet (Chen et al., 2021), were trained and evaluated on the Pediatric-N CXR dataset to record baseline performance. The Pediatric-N dataset was constructed by adding to the private pediatric CXR collection from (i), the pediatric CXRs that were removed from the Shenzhen (Jaeger et al., 2014), Montgomery (Jaeger et al., 2014), and Japanese Society of Radiological Technology (JSRT) (Oda et al., 2009) CXR datasets.

(ii)    The best-performing model architecture from (ii) was trained on the Shenzhen-N, Montgomery-N, and JSRT-N CXR datasets in isolation and their combinations. The term $N$ denotes the datasets after the removal of the pediatric CXRs.

(iii)   The predictions of the top-K (K = 2, 3) adult models from (iii) were aggregated through bitwise operations and a stacked ensemble.

(iv)    We further construct an ensemble of stacked ensembles as follows: The predictions of the stacked ensembles constructed using the top-2 and top-3 adult models from (iv), were aggregated using bitwise operations.

(v)     The performance of the models from (ii) to (v) was evaluated using novel evaluation metrics, which are the mean lung contour distance (MLCD), average hash score (AHS), in addition to the widely used multi-scale structural similarity index measure (MS-SSIM), the intersection of union (IoU), Dice score, 95% Hausdorff distance (HD95), and average symmetric surface distance (ASSD).

(vi)    We report statistical significance in the Dice score achieved by the models from (ii) to (v) using 95% binomial confidence intervals (CIs) measured using the Clopper-Pearson Exact method.

(vii)    Through extensive empirical analyses, we observed significant improvements in performance and cross-domain generalization through this systematic approach.

The remaining parts of this study are organized as follows: We elaborate on the datasets, models, training parameters, loss functions, evaluation metrics, and statistical analysis in Section 3. Section 4 discusses the experimental results. Finally, the conclusions of this study and the scope for future research are discussed in Section 5.

# 3. MATERIALS AND METHODS

## 3.1. Datasets

The following datasets were used in this study:

Shenzhen CXR dataset: This publicly available collection includes 662 deidentified frontal CXR projections obtained from the Shenzhen No. 3 hospital in Shenzhen, China. 326 images show normal lungs, and 336 images exhibit pulmonary abnormalities consistent with TB along with radiological findings, age, and gender. The pixel-wise ground truth (GT) annotations for TB-consistent manifestations are made publicly available (Yang et al., 2022). In (Stirenko et al., 2018), the authors released the GT masks annotating lung ROIs for a selection of 566 CXRs. Of these, we excluded 43 CXRs that erroneously included the heart regions in the GT lung masks. The remaining 523 CXRs included 24 pediatric CXRs of various age groups, the details are given in the first row of Table 1.

Montgomery CXR dataset: This publicly available collection includes 138 frontal CXR images collected through the TB screening program in Montgomery County, Maryland, USA. There are 58 images with abnormalities consistent with TB and 80 images showing normal lungs. The dataset includes de-identified CXRs and their associated radiological findings, GT masks for lung ROIs, and the metadata including age and gender. It includes 17 pediatric CXRs belonging to various age groups which are detailed in the second row of Table 1.

Table 1 – Characteristics of CXR datasets.

| Datasets | Age groups | | | |
|---|---|---|---|---|
| | 1 day to < 24 months | 24 months to < 11 years | 11 years to < 18 years | >=18 years |
| Shenzhen | 5 | 14 | 5 | 499 |
| Montgomery | 0 | 7 | 10 | 121 |
| JSRT | 0 | 0 | 2 | 245 |
| Pediatric | 52 | 67 | 42 | 0 |
| Pediatric-N | 57 | 88 | 59 | 0 |

JSRT CXR dataset: This publicly available collection includes 247 frontal CXR images and their associated masks for the lungs, clavicles, and heart regions. The data was collected as a part of the research activities of the academic committee supported by the Japanese Radiological Society (JRS). Of the 247 CXRs, 100 CXRs exhibit malignant nodules, 54 CXRs contain benign nodules, and 93 CXRs show normal lungs. The related metadata consisting of age, gender, radiological diagnosis, and location and degree of subtlety of the nodules are publicly available. It includes 2 pediatric CXRs, the details are summarized in the third row of Table 1.

Pediatric CXR dataset: This is a private collection of 161 frontal CXRs and their associated lung masks (Candemir et al., 2015). The radiographic projections are captured from the children of various age groups for a lung segmentation task. The CXRs are de-identified and their use is exempted by the Institutional Review Board (IRB) of the National Institutes of Health (NIH). The CXRs are further divided into three groups, viz., group 1 (1 day to < 24 months), group 2 (24 months to < 11 years), and group 3 (11 years to < 18 years) based on the lung developmental stages from infancy to adulthood as discussed in the literature (Candemir et al., 2015), and is shown in the fourth row of Table 1.

The CXRs and masks in each dataset were divided at the patient level into 70% for training, 10% for validation, and 20% for hold-out testing. The number of images across the train, validation, and test splits is shown in Table 2. Here, Shenzhen-N, Montgomery-N, and JSRT-N denote Shenzhen, Montgomery, and JSRT CXR datasets after the removal of pediatric CXRs. The pediatric CXRs removed from the Shenzhen, Montgomery, and JSRT CXR datasets were added to the appropriate age groups in the Pediatric CXR dataset to form the Pediatric-N dataset. The details of the number of CXRs in the Pediatric-N dataset are given in the last row of Table 1. The hold-out test set constructed from the pediatric-N dataset was further divided into three groups. The terms P1, P2, and P3 denote the number of test images in pediatric group 1, group 2, and group 3, respectively. The images and their associated masks in the train, validation and hold-out test sets were rescaled to 224 × 224 pixels to reduce computational complexity and were normalized to the range [0,1]. We henceforth use the acronyms P, S, M, and J to denote the Pediatric-N, Shenzhen-N, Montgomery-N, and JSRT-N CXR datasets, respectively.

Table 2 – Datasets with their train, validation, and test splits. The term NA denotes "Not Available".

| Dataset | Train | Validation | Test-1 | Test-2 | Test-3 |
|---|---|---|---|---|---|
| Pediatric-N (P) | 144 | 21 | 11 (P1) | 17 (P2) | 11 (P3) |
| Shenzhen-N (S) | 350 | 50 | 99 | NA | NA |
| Montgomery-N (M) | 85 | 13 | 23 | NA | NA |
| JSRT-N (J) | 172 | 25 | 48 | NA | NA |

*3.2. Model architectures*

Fig. 3 illustrates the general architecture of the UNet and TransUNet models. UNet is a fully convolutional U-shaped network that performs pixel-wise classification using an encoder/contracting path followed by a decoder/expanding path. The encoder uses convolutional and max-pooling layers to learn input semantics, i.e., complex information in the input data, by expanding the receptive field while condensing the input information to a latent space feature representation. The decoder has the same architecture as at each level of the encoder but performs the opposite function by (i) using un-padded or padded transposed convolutions/deconvolutions to correspondingly reduce the number of channels, (ii) up-sample the condensed information at the bottom of the encoder back to the original input shape and (iii) localize the ROIs in the input. Skip connections are used at each level where the feature maps of the encoder are concatenated to the output of the transposed convolutions at the same level in the decoder to facilitate precise ROI localization. The final convolutional layer in the decoder predicts the grayscale masks.

TransUNet model has a U-shaped architecture like the UNet model. However, it uses a hybrid architecture composed of CNN and Transformers in the encoding path. The input image is split into multiple patches (n = 16). A CNN model extracts features from these patches. The learned feature maps are tokenized into a two-dimensional embedding by feeding into the Vision Transformer (ViT) (Chen et al., 2021) that uses multi-head self-attention (number of heads = 12) and fully-connected multi-layer perceptron (dimension = 3072) modules to encode short- and long-range dependencies and learn hidden feature representations. The learned representations are reshaped and fed into a cascading up sampler/decoder that uses four 2 × Upsampling blocks to correspondingly upscale the condensed information at the bottom of the encoder to achieve the original input resolution. Skip connections are used for concatenating the extracted features from these patches at the encoder to the decoder at any given level. The final convolutional layer predicts the grayscale masks.

In this study, we used segmentation models consisting of (i) TransUNet (Chen et al., 2021), (ii) the conventional UNet (Ronneberger et al., 2015), and (iii) the UNet variants with several ImageNet-pre-trained encoder backbones, viz., VGG-16, VGG-19, Inception-V3, ResNet-50, DenseNet-121, and EfficientNet-B0, and their corresponding decoders constructed using a publicly available Python library (Iakubovskii, 2019). For the TransUNet, we used the ImageNet-pre-trained ResNet-50 and ViT-B/16 in the encoder backbone as discussed in the literature (Chen et al., 2021). The masks predicted by the UNet and TransUNet models were thresholded and binarized to separate the lung ROI/foreground pixels from the non-lung regions/background.

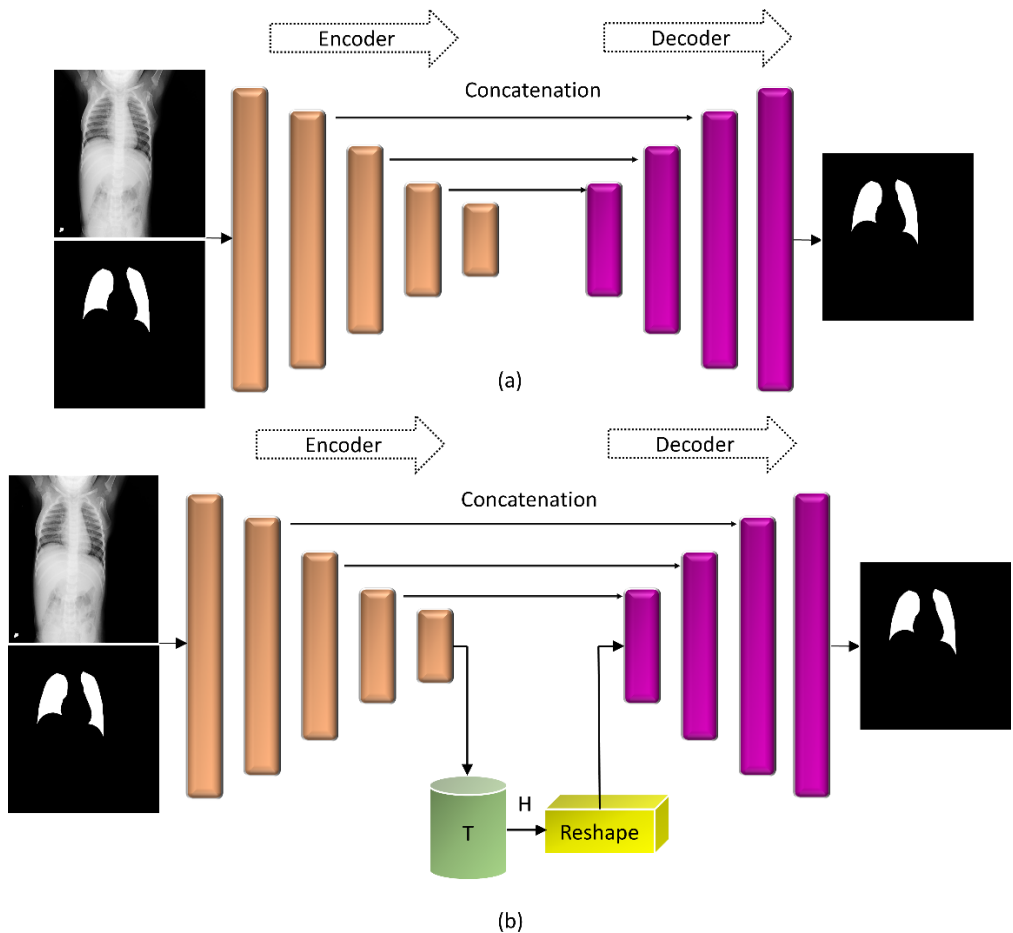

Fig. 3 – The general architecture of the (a) UNet and (b) TransUNet model. Here, T denotes the Transformer and H denotes the learned hidden representation.

### 3.3. Ensemble learning

Our previous findings (Rajaraman et al., 2022) showed that applying bitwise operations to the top-3 model predictions improved performance in a TB-consistent lesion segmentation task. Therefore, we combined the binarized predictions of the top-K (K = 2, 3) adult models using the bitwise-OR and bitwise-AND operations to evaluate performance using the pediatric test groups P1, P2, and P3.

A stacked ensemble combines the predictions from models using a second-level meta-learner that (i) harnesses the individual models' capabilities by learning a diverse feature space, (ii) makes final predictions with reduced prediction variance, and (iii) improves generalization compared to any individual model (Wolpert, 1992). We constructed a stacked ensemble using the top-K (K = 2, 3) adult models as follows: (i) The top-K adult models were instantiated with their CXR modality-specific pretrained weights. (ii) The trainable layers of these models were frozen. (iii) The feature maps extracted from the penultimate layer of these models were concatenated. (iv) A meta-learner, consisting of a fully convolutional network with six convolutional layers learned to output the final prediction. We used 512, 256, 128, 64, and 32 convolutional filters of size 3×3 with ReLU activations, respectively, in the $1^{st}$ through $5^{th}$ convolutional layers. The final convolutional layer of the meta-learner with a single filter of dimension 1×1 and sigmoidal activation predicted the grayscale mask.

We further constructed an ensemble of stacked ensembles to observe improvements in segmentation performance. Here, we performed bitwise-OR and bitwise-AND of the predictions of the stacked ensembles constructed using the top-2 and top-3 adult models, respectively.

### 3.4. Training parameters, loss function, and evaluation metrics

We used a modified Focal-Tversky loss function with an additional boundary uncertainty component (Yeung et al., 2021) for our lung segmentation task. The literature (Rajaraman et al., 2022) demonstrated superior segmentation performance in a fine-grained TB-consistent region segmentation task using this loss function. We used the Adam optimizer with an initial learning rate of $1 \times 10^{-3}$, a batch size of 16, and 128 epochs to train the models. The learning rate was reduced by a factor of 2 if the validation performance did not improve after 10 epochs. All experiments were conducted in an Ubuntu Linux system containing an Nvidia GeForce GTX1080Ti GPU with CUDA dependencies and Tensorflow Keras v2.6.

We used pixel-based, image-based, and distance-based evaluation metrics in this study. The pixel-based evaluation aims at evaluating the performance of the model by assigning a semantic class to every individual pixel. A study of the literature (Candemir et al., 2016; Hesamian et al., 2019; Zamzmi et al., 2022) reveals that pixel-based metrics such as IoU and Dice scores are widely used to evaluate segmentation models.

Intersection of Union (IoU): The IoU metric is a widely used pixel-based metric to evaluate segmentation performance. Here, TP, FP, and FN denote the true positives, false positives, and false negatives respectively. Given an IoU threshold, the prediction is considered true positive if the overlapping with the GT exceeds this threshold. False negatives denote GT having no associated predictions and false positives denote predictions with no associated GT.

$$IoU = \frac{TP}{TP + FP + FN} \tag{1}$$

Dice: The Dice score is another widely used pixel-based evaluation metric in semantic segmentation. The IoU and Dice range from [0, 1]. Values closer to 1 denote increasing similarity of the predictions to the GT.

$$Dice\ score = \frac{2TP}{2TP + FP + FN} \tag{2}$$

Evaluating only using pixel-based metrics may ignore the dependencies among the image pixels. Such an approach may not help fully explore a model's segmentation potential. A structural similarity index measure (SSIM)-based loss function was proposed in the literature (S. Zhao et al., 2019) to study its impact on learning the structural details. The authors demonstrated improved performance with SSIM loss compared to the conventional cross-entropy loss.

Motivated by (S. Zhao et al., 2019), we used MS-SSIM, an extension of SSIM, as an image-based evaluation metric in our study. The MS-SSIM is more robust than the SSIM as it measures the similarity at multiple input scales which helps to pay attention to the pixel dependencies at multiple scales between corresponding regions in the GT and predictions. The MS-SSIM of a pair of images ($a$, $b$) is given by a multiplicative combination of the luminance ($l$), contrast ($c$), and structural ($s$) terms at multiple scales.

$$SSIM\ (a,b) = [l(a,b)]^{\alpha} \cdot [c(a,b)]^{\beta} \cdot [s(s,b)]^{\gamma} \tag{3}$$

$$MS - SSIM\ (a,b) = [l_m(a,b)]^{\alpha_M} \cdot \prod_{j=1}^{M} [c_j(a,b)]^{\beta_j} \cdot [s_j(a,b)]^{\gamma_j} \tag{4}$$

Here, $\alpha, \beta, \gamma$ denote the weight given to each term and $M$ corresponds to the lowest image resolution i.e.,

the highest number of down-sampling operations performed to reduce image resolution. The value of $j = 1$ denotes the original input resolution. The input was down-sampled by a factor of 2 with each scaling. The quality map would be identical in size to the corresponding scaled version of the image. We fixed the value of $M = 5$. The MS-SSIM value ranges from [0, 1].

We used the quality maps computed at multiple scales to visualize the segmentation quality and analyze the similarity between the GT and predicted masks. Using the Jet colormap array, an analysis of the quality maps revealed high activations in red color for regions of superior prediction quality. Here, the MS-SSIM values tend to achieve a value close to or equal to 1. Blue region activations denoted poor prediction quality where the MS-SSIM tends to achieve a value close to or equal to 0.

We propose a novel, image-based evaluation metric, called the average hash score (AHS) that uses hashing algorithms in spatial and frequency domains to analyze the similarity between the GT and predicted masks. Image hashing is used in applications including analyzing image similarity in spatial and frequency domains, data structures, and error detection, among others (Ou & Rhee, 2010). A hash function maps image data of arbitrary size to fixed-size values called hashes. Similar GT and predicted masks would have the same hash values. We used the spatial domain hashing methods consisting of average hashing (aHash), difference hashing (dHash), and the frequency domain hashing methods consisting of perceptual hashing (pHash) and wavelet hashing (wHash). The details of these hashing methods can be found in the literature (Ou & Rhee, 2010). The AHS is given by the average of the aforementioned hashing methods.

$$Average\ hash\ score\ (AHS) = \frac{1}{4} * (aHash + \ dHash + pHash + wHash) \quad (5)$$

A value closer to 0 denotes increasing similarity between the GT and predicted masks, and a value closer to 1 denotes increasing dissimilarity. The AHS score is a robust evaluation metric since it evaluates similarity in both spatial and frequency domains.

We propose a novel, distance-based evaluation metric, called the Mean Lung Contour Distance (MLCD) to analyze the similarity between the GT and predicted mask contours and quantify the distance between these contours. A contour is defined as a curve joining the continuous points that have similar intensity along the image boundary. Contours play a vital role in several applications including shape analysis, segmentation, and object detection, among others. Fig. 4 illustrates the workflow for computing the MLCD. First, a Canny edge detector was applied to extract the contours of the binarized GT mask. Then, a distance map was generated based on the GT contour. Each pixel in the distance map was assigned the Euclidean distance between it and the nearest non-zero pixel (i.e., pixels on the GT contour). Next, the Canny edge detector was applied to extract the contours of the predicted mask. The distance between the contours of the GT and predicted masks was measured as the sum of the distance map values along the predicted mask contour. Finally, the MLCD metric was computed as the mean of the distance values.

$$MLCD = \frac{1}{N} \sum_{i \in C_{pred}} D_i \tag{6}$$

Here, $C_{pred}$ denotes the predicted mask contour, $N$ denotes the number of pixels along the predicted mask contour, and $D_i$ denotes the pixel value in the distance map along the predicted mask contour.

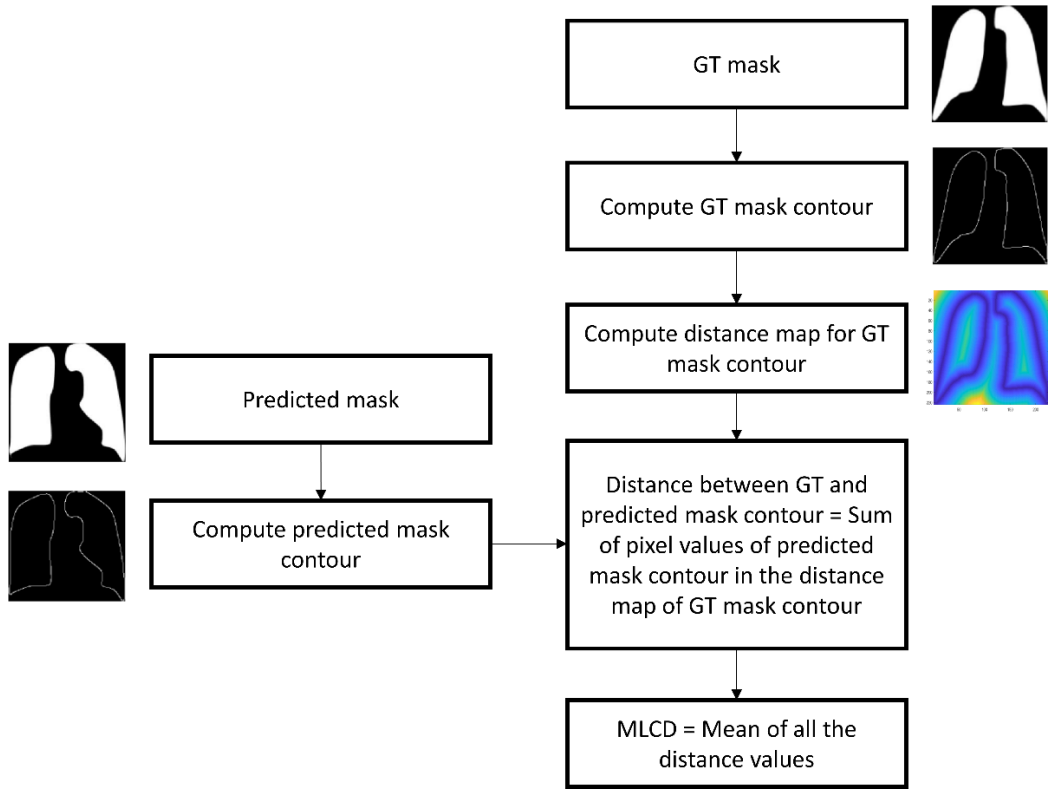

Fig. 4. Workflow for MLCD computation using a pair of GT and predicted masks.

We measured the 95% Hausdorff distance (HD95) and the average symmetric surface distance (ASSD)

to evaluate segmentation performance. The HD metric is a measure of the difference between two sets in a metric space. This measure is useful in applications such as image segmentation, where the sets would represent sets of points corresponding to features in the images. Given two sets *A* and *B* in a metric space *(X, d)*, the HD between *A* and *B*, denoted as $d_{H(A,B)}$, is defined as the maximum of the minimum distances between each point in *A* and *B*.

$$d_{H(A,B)} = \max\{ h(A,B), h(B,A)\} \tag{7}$$

$$H(A,B) = \max a \text{ in } A\{\min b \text{ in } B\{d(a,b)\} \tag{8}$$

Here, $H(A,B)$ denotes the maximum over all the points in A of the minimum distance between that point and any point in B. The HD95 metric refers to the value below which 95% of the HDs between all possible pairs of sets fall. The HD95 metric is a robust estimate of maximum error compared to HD, which is highly sensitive to small outliers (Aydin et al., 2021). A smaller value for HD95 denotes a higher similarity between the GT and the predicted masks.

The average symmetric surface distance (ASSD) is commonly used in image segmentation applications to quantify the similarity between GT and predicted masks. It is a measure of the difference between two shapes in a two-dimensional space. Given two shapes *S1* and *S2*, the ASSD is defined as the average of the Euclidean distances between corresponding points on the two shapes. The term "symmetric" refers to the fact that the distance is computed in both directions, and the average is taken over both directions. The formal definition of the ASSD between two surfaces *S1* and *S2* can be expressed as follows.

$$ASSD(S1,S2) = \left(\frac{1}{N}\right) * \left( \Sigma_{\{i=1\}}^{N} d(p1_i, p2_i) + \Sigma_{\{i=1\}}^{N} d(p2_i, p1_i)\right) \tag{9}$$

Here, *N* denotes the number of corresponding points on the two surfaces, $p1_i$ is a point on surface *S1*, $p2_i$ is the corresponding point on surface *S2*, and $d(p1_i, p2_i)$ is the Euclidean distance between points $p1_i$ and $p2_i$. A smaller value for ASSD indicates that the two surfaces are more similar, while a larger ASSD indicates that the two surfaces are more different. Table 3 summarizes the properties of the several pixel-based, image-based, and distance-based evaluation metrics used in this study.

Table 3 – Characteristics of the evaluation metrics

| Evaluation metric | Evaluation method | Similarity | Visualization |
|---|---|---|---|
| IoU | Pixel-based | No | No |
| Dice Score | Pixel-based | No | No |
| MS-SSIM | Image-based | Yes | Yes (quality maps) |
| AHS | Image-based | Yes | No |
| MLCD | Distance-based | Yes | No |
| HD95 | Distance-based | Yes | No |
| ASSD | Distance-based | Yes | No |

*3.5. Statistical analysis*

We report statistical significance for the Dice score using the 95% binomial CIs, measured using the Clopper–Pearson Exact method to discriminate model performance. We followed the guidelines in (Altman & Bland, 2011) to measure the probability value (*p*-value) from the CIs.

## 4. RESULTS AND DISCUSSION

Table 4 shows the performance achieved by the UNet and TransUNet models with the pediatric test groups P1, P2, and P3. Note that the UNet model with the Inception-V3 encoder backbone demonstrated superior performance in terms of all metrics across the pediatric test groups. A lower value for the MLCD, HD95, and ASSD, and a higher value for the IoU, Dice scores, MS-SSIM, and AHS, underscore the fact that the masks predicted by the Inception-V3-UNet were markedly similar to the GT. The performance of the Inception-V3-UNet was followed by the ResNet-50-UNet and TransUNet models. We observed that the performance of the Inception-V3-UNet model was significantly superior compared to the VGG-16-UNet, VGG-19-UNet, DenseNet-121-UNet, and EfficientNet-B0-UNet models across the pediatric test groups ($p < 0.05$). Therefore, we used this model architecture for further experimental evaluations.

Table 4 – Test performance achieved by the various segmentation models. Bold numerical values denote superior performance. The 95% CIs for the Dice score was measured using the Clopper–Pearson exact method. The * denotes statistical significance ($p < 0.05$).

| Test groups | Models | IoU | Dice (CI) | MS-SSIM | MLCD | AHS | HD95 | ASSD |
|---|---|---|---|---|---|---|---|---|
| P1 (1 day to < 24 months) | UNet (Ronneberger et al., 2015) | 0.5830 | 0.7366 (0.4762,0.9970) | 0.8522 | 16.3217 | 8.5909 | 66.7308 | 51.0221 |
| | VGG-16-UNet | 0.8057 | 0.8924 (0.7092,1.0) | 0.8987 | 14.0836 | 9.8182 | 62.6655 | 108.0936 |
| | VGG-19-UNet | 0.4222 | 0.5937 (0.3034,0.8840) | 0.7040 | 17.7648 | 14.5000 | 71.4703 | 83.5034 |
| | Inception-V3-UNet | **0.8637** | **0.9268 (0.7728,1.0)\*** | **0.9745** | **3.4513** | **7.0228** | **39.3002** | **41.2564** |
| | ResNet-50-UNet | 0.8589 | 0.9241 (0.7675,1.0) | 0.9714 | 5.9878 | 8.4546 | 41.0295 | 40.2435 |
| | DenseNet-121-UNet | 0.2978 | 0.4589 (0.1644,0.7534) | 0.4899 | 15.6464 | 17.5000 | 72.4469 | 62.7347 |
| | EfficientNet-B0-UNet | 0.8144 | 0.8977 (0.7186,1.0) | 0.9503 | 9.6525 | 7.9773 | 60.0399 | 61.1625 |
| | TransUNet | 0.8506 | 0.9193 (0.7583,1.0) | 0.9644 | 3.8082 | 7.9319 | 56.8448 | 42.1224 |
| P2 (24 months to < 11 years) | UNet (Ronneberger et al., 2015) | 0.7582 | 0.8625 (0.6987,1.0) | 0.8165 | 11.5392 | 8.2059 | 52.2579 | 62.9313 |
| | VGG-16-UNet | 0.8341 | 0.9095 (0.7731,1.0) | 0.8722 | 8.3727 | 6.9853 | 47.1770 | 71.3197 |
| | VGG-19-UNet | 0.6358 | 0.7774 (0.5796,0.9752) | 0.7052 | 14.9264 | 11.8383 | 59.0714 | 75.5571 |
| | Inception-V3-UNet | **0.9004** | **0.9476 (0.8416,1.0)\*** | **0.9360** | **4.5234** | **4.8677** | **29.1362** | **52.9087** |
| | ResNet-50-UNet | 0.8888 | 0.9411 (0.8291,1.0) | 0.9302 | 5.3729 | 5.6030 | 35.5154 | 57.4032 |
| | DenseNet-121-UNet | 0.4456 | 0.6165 (0.3853,0.8477) | 0.4494 | 13.3331 | 16.9853 | 65.0131 | 73.8481 |
| | EfficientNet-B0-UNet | 0.8731 | 0.9322 (0.8126,1.0) | 0.9124 | 6.6996 | 5.4853 | 38.3640 | 92.5570 |
| | TransUNet | 0.8948 | 0.9445 (0.8356,1.0) | 0.9343 | 5.1735 | 5.5148 | 32.9401 | 57.1883 |
| P3 (11 years to < 18 years) | UNet (Ronneberger et al., 2015) | 0.7023 | 0.8251 (0.6006,1.0) | 0.7703 | 12.0381 | 8.8409 | 37.8993 | 70.6874 |
| | VGG-16-UNet | 0.7679 | 0.8687 (0.6691,1.0) | 0.8318 | 10.3997 | 5.1591 | 37.3493 | 64.3771 |
| | VGG-19-UNet | 0.6213 | 0.7664 (0.5163,1.0) | 0.6538 | 15.3366 | 11.2046 | 59.7629 | 85.4540 |
| | Inception-V3-UNet | **0.9330** | **0.9654 (0.8573,1.0)\*** | **0.9390** | **4.1418** | **2.4546** | **14.2044** | **61.7601** |
| | ResNet-50-UNet | 0.9269 | 0.9621 (0.8492,1.0) | 0.9364 | 4.8088 | 2.5455 | 31.9757 | 66.1349 |
| | DenseNet-121-UNet | 0.4807 | 0.6493 (0.3672,0.9314) | 0.4254 | 13.5877 | 18.1591 | 60.1191 | 96.3356 |
| | EfficientNet-B0-UNet | 0.9140 | 0.9551 (0.8327,1.0) | 0.9215 | 4.0578 | 2.6137 | 34.1645 | 67.6853 |
| | TransUNet | 0.9224 | 0.9597 (0.8434,1.0) | 0.9318 | 4.4899 | 2.5910 | 32.8182 | 66.7702 |

Table 5 shows the performance achieved with the pediatric test groups using the Inception-V3-UNet model trained with the P, and the S, M, and J CXR datasets, in isolation as well as combinations. Fig. 5 shows the model predictions for sample CXRs from the three pediatric test groups. The results for the Inception-V3-UNet model trained with the P dataset come from Table 4. From Table 5 and Fig. 5, we note that for the pediatric test group P1, the Inception-V3-UNet model trained on the S, M, and J adult CXR datasets in isolation and combination could not outperform that trained on the P CXR dataset. This may be because of significant differences in the lung shapes of infants in the age span of 1 day to 2 years as compared to the adult lungs as discussed in the literature (Arthur, 2000; Candemir et al., 2015). This is further confirmed through a higher value for the MLCD, HD95, and ASSD metrics demonstrated by the adult models, unlike the pediatric model, due to the increasing contour distances between the predicted and GT masks. For the P2 test group, we observed that the Inception-V3-UNet model trained on the collection of S, M, and J adult CXR datasets outperformed that trained on the P CXR dataset in terms of all metrics.

Table 5 – Performance achieved by the Inception-V3-UNet model trained on the P CXR dataset, and the S, M, and J adult CXR datasets, in isolation as well as combinations, and tested with the pediatric test groups P1, P2, and P3. Bold numerical values denote superior performance achieved within each pediatric test group. The * denotes statistical significance ($p < 0.05$).

| Test groups | Training dataset(s) | IoU | Dice (CI) | MS-SSIM | MLCD | AHS | HD95 | ASSD | Rank (Dice) |
|---|---|---|---|---|---|---|---|---|---|
| P1 | P | **0.8637** | **0.9268 (0.7728,1.0)*** | **0.9745** | **3.4513** | 7.0228 | 39.3002 | 41.2564 | **1** |
| | S | 0.7858 | 0.8801 (0.6881,1.0) | 0.9298 | 11.9667 | 7.9091 | 56.0733 | 53.1736 | 2 |
| | M | 0.5645 | 0.7217 (0.4568,0.9866) | 0.7515 | 16.0476 | 13.0682 | 71.2458 | 119.5940 | |
| | J | 0.4656 | 0.6353 (0.3508,0.9198) | 0.6552 | 16.0949 | 15.3410 | 79.7541 | 133.8766 | |
| | S & M | 0.7661 | 0.8675 (0.6671,1.0) | 0.9195 | 14.3176 | 7.8864 | 75.3041 | 60.6665 | 4 |
| | S & J | 0.7795 | 0.8761 (0.6813,1.0) | 0.9255 | 12.5665 | 6.3864 | 57.5609 | 56.3334 | 3 |
| | M & J | 0.5740 | 0.7293 (0.4667,0.9919) | 0.7839 | 16.4522 | 12.1819 | 76.8209 | 108.8101 | |
| | S, M, & J | 0.6535 | 0.7904 (0.5498,1.0) | 0.9151 | 15.6954 | 7.5910 | 75.4847 | 66.3901 | |
| P2 | P | 0.9004 | 0.9476 (0.8416,1.0) | 0.9360 | 4.5234 | 4.8677 | 29.1362 | 52.9087 | |
| | S | 0.9055 | 0.9504 (0.8471,1.0) | 0.9361 | 5.2153 | 4.8089 | 38.0552 | 58.5496 | 3 |
| | M | 0.8517 | 0.9199 (0.7908,1.0) | 0.8369 | 12.3820 | 6.7648 | 53.9304 | 98.5443 | |
| | J | 0.7429 | 0.8525 (0.6839,1.0) | 0.7411 | 14.2227 | 9.3530 | 63.8007 | 119.3509 | |
| | S & M | 0.8960 | 0.9452 (0.837,1.0) | 0.9319 | 5.9698 | 4.3971 | 42.2812 | 62.2580 | |
| | S & J | 0.9080 | 0.9518 (0.8499,1.0) | 0.9460 | 5.1572 | 4.2795 | 26.6672 | 52.0278 | 2 |
| | M & J | 0.8341 | 0.9096 (0.7732,1.0) | 0.8581 | 12.3950 | 6.3971 | 57.8081 | 85.5697 | |
| | S, M, & J | **0.9111** | **0.9535 (0.8534,1.0)*** | **0.9494** | **3.4200** | **3.3530** | **10.4097** | **51.3569** | **1** |
| P3 | P | 0.9330 | 0.9654 (0.8573,1.0) | 0.9390 | 4.1418 | 2.4546 | 16.8817 | 60.2842 | |
| | S | 0.9336 | 0.9657 (0.8581,1.0) | 0.9482 | 2.6479 | 2.7955 | 18.1527 | 61.7601 | |
| | M | 0.9430 | 0.9707 (0.871,1.0) | 0.9498 | 2.6220 | 1.8864 | 14.2044 | 60.4817 | 3 |
| | J | 0.8751 | 0.9334 (0.786,1.0) | 0.8328 | 13.0810 | 5.0000 | 33.9239 | 87.2765 | |

| | | | | | | | | |
|---|---|---|---|---|---|---|---|---|
| S & M | 0.9529 | 0.9759 (0.8852,1.0) | 0.9676 | 1.4804 | 1.7046 | 2.9931 | 55.9776 | 2 |
| S & J | 0.9118 | 0.9539 (0.8299,1.0) | 0.9419 | 3.3546 | 2.6819 | 20.3217 | 70.1243 | |
| M & J | 0.9395 | 0.9688 (0.8660,1.0) | 0.9422 | 2.9512 | 1.9546 | 16.2292 | 66.9049 | |
| S, M, & J | **0.9544** | **0.9767 (0.8875,1.0)\*** | **0.9686** | **1.4398** | **1.4546** | **2.9150** | **61.2869** | **1** |

A similar increase in performance with the P3 test group is observed with multi-institutional data training using the collection of S, M, and J adult CXR datasets compared to that trained on the P CXR dataset.

Fig. 6 shows the MS-SSIM quality maps for sample CXRs from the pediatric test groups P1, P2, and P3. High activations in red regions indicate superior prediction quality, i.e., high similarity to the GT masks, with MS-SSIM values that are close to 1. In contrast, blue region activations indicate poor prediction quality. The quality maps (Fig. 6(a) bottom) show poorer quality predictions for a sample CXR from the P1 test group, by the model trained on the S adult CXR dataset compared to that trained on the P CXR dataset (Fig. 6(a) top). But for the P2 test group, we could observe increased red activations for the prediction by the Inception-V3-UNet model trained on the collection of S, M, and J adult CXR datasets (Fig. 6(b) bottom) compared to that trained on the P CXR dataset (Fig. 6(b) top). These findings are consistent with those reported in the literature (Therrien & Doyle, 2018) which recommended increasing training data variability through acquisitions from multiple sites and testing with other external data. We believe that our multi-institutional training data training might have introduced sufficient variability into the training process and resulted in improved performance.

The MS-SSIM quality maps (Fig. 6(c) bottom) reinforce these observations showing markedly improved prediction quality by the Inception-V3-UNet model trained on the collection of S, M, and J adult CXR datasets compared to that trained on the P CXR dataset (Fig. 6(c) top). Our findings echo the observations in the literature (Candemir et al., 2015) where lung contours of P3 are of subjects that are reasonably mature in age and therefore similar to the adult lungs.

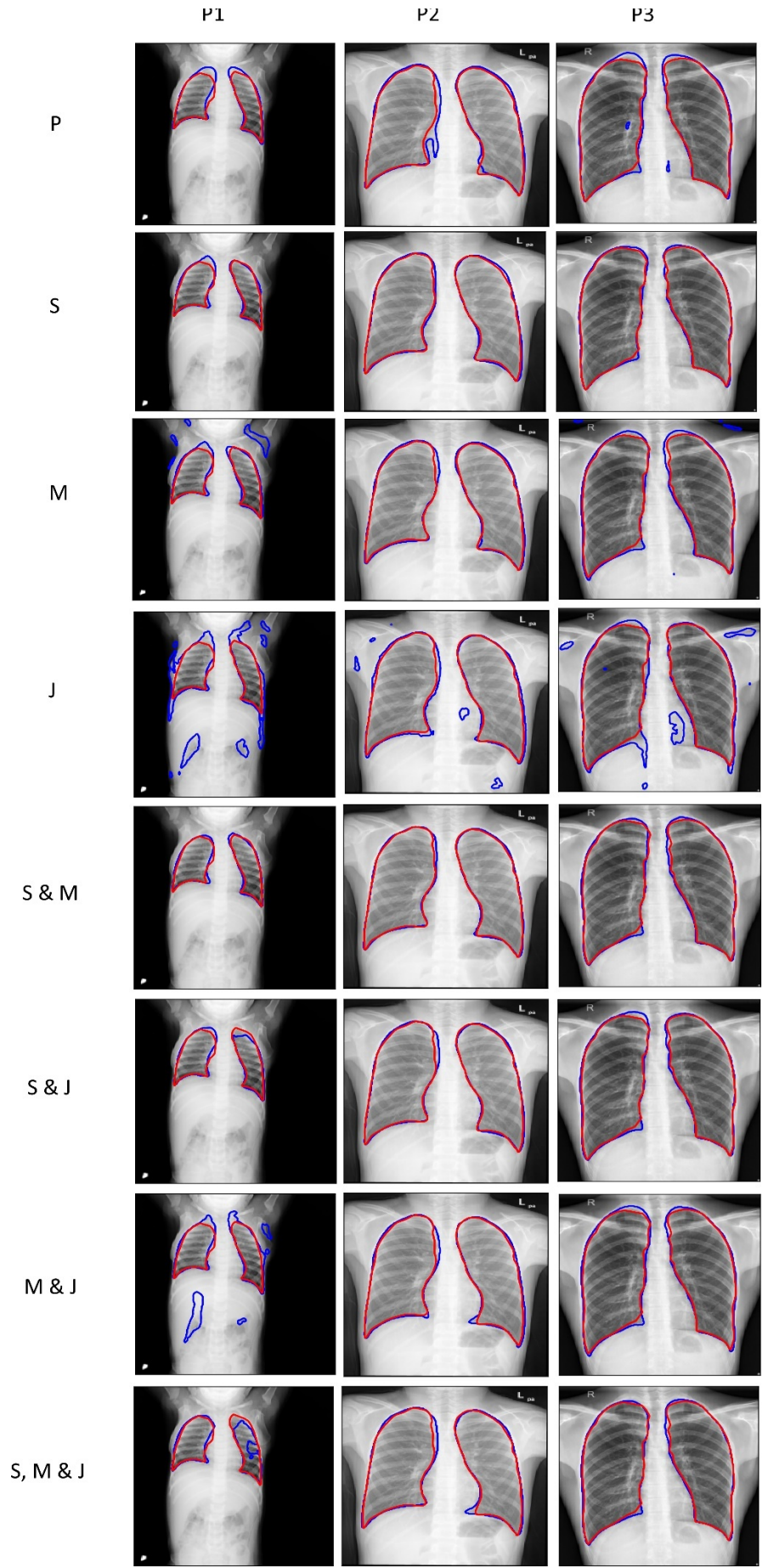

Fig. 5 – Visualizing segmentation predictions of the Inception-V3-UNet trained on various datasets for sample CXR instances from P1 (age = 11 months), P2 (age = 8 years ), and P3 (age = 14 years) test groups. The red and blue contours denote GT and predictions, respectively.

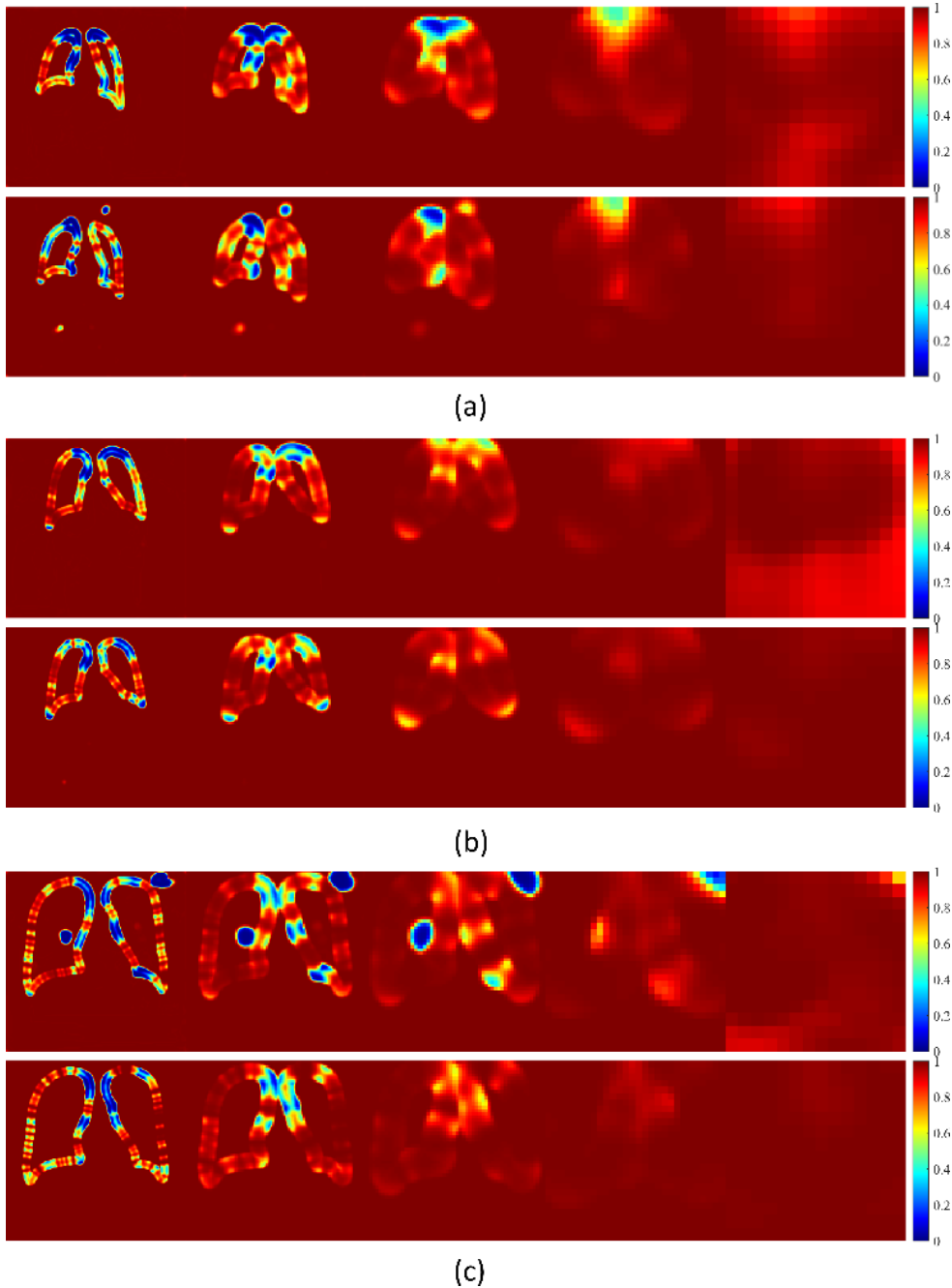

(a)

(b)

(c)

Fig. 6 – MS-SSIM quality maps at five different scales. (a) The prediction of the Inception-V3-UNet model trained with the P CXR dataset (top) and S adult CXR dataset (bottom) for an instance of CXR (age = 10 months) from pediatric test group P1. (b) The prediction of the Inception-V3-UNet model trained with the P CXR dataset (top) and the combined S, M, and J adult CXR datasets (bottom) for an instance of CXR (age = 3 years) from pediatric test group P2. (c) The prediction of the Inception-V3-

UNet model trained with the P CXR dataset (top) and the combined S, M, and J adult CXR datasets (bottom) for an instance of CXR (age = 15 years) from pediatric test group P3.

We performed bitwise-OR and bitwise-AND operations using the predictions of the top-K (K = 2, 3) adult models for each pediatric test group P1, P2, and P3. The models were ranked based on the Dice metric (Table 5). Table 6 shows the performance achieved in this regard.

Table 6 – Test performance achieved through bitwise operations of the predictions of the top-K (K = 2, 3) adult models. The baseline denotes the test performance achieved by the Inception-V3-UNet model trained on the P CXR dataset. Bold numerical values denote superior performance in each test group. The * denotes statistical significance ($p < 0.05$).

| Test groups | Method | Models | IoU | Dice (CI) | MS-SSIM | MLCD | AHS | HD95 | ASSD |
|---|---|---|---|---|---|---|---|---|---|
| P1 | Baseline | P | **0.8637** | **0.9268 (0.7728,1.0)*** | **0.9745** | **3.4513** | **7.0228** | **39.3002** | **41.2564** |
| | Bitwise-AND | Top-2 | 0.7911 | 0.8834 (0.6937,1.0) | 0.9464 | 9.6752 | 7.5910 | 42.9932 | 54.1293 |
| | | Top-3 | 0.7538 | 0.8596 (0.6542,1.0) | 0.9054 | 13.8131 | 8.4091 | 52.6717 | 72.6014 |
| | Bitwise-OR | Top-2 | 0.7538 | 0.8596 (0.6542,1.0) | 0.9054 | 13.8131 | 8.4091 | 52.6717 | 72.6014 |
| | | Top-3 | 0.7538 | 0.8596 (0.6542,1.0) | 0.9054 | 13.8131 | 8.4091 | 52.6717 | 72.6014 |
| P2 | Baseline | P | 0.9004 | 0.9476 (0.8416,1.0) | 0.9360 | 4.5234 | 4.8677 | 29.1362 | 52.9087 |
| | Bitwise-AND | Top-2 | **0.9183** | **0.9574 (0.8613,1.0)** | **0.9562** | **2.7597** | **3.3900** | **20.4014** | **43.7382** |
| | | Top-3 | 0.9036 | 0.9494 (0.8452,1.0) | 0.9393 | 5.6449 | 4.1700 | 25.1912 | 47.4764 |
| | Bitwise-OR | Top-2 | 0.9036 | 0.9494 (0.8452,1.0) | 0.9393 | 5.6449 | 4.1700 | 25.1912 | 47.4764 |
| | | Top-3 | 0.9036 | 0.9494 (0.8452,1.0) | 0.9393 | 5.6449 | 4.1700 | 25.1912 | 47.4764 |
| P3 | Baseline | P | 0.9330 | 0.9654 (0.8573,1.0) | 0.9390 | 4.1418 | 2.4546 | 16.8817 | 60.2842 |
| | Bitwise-AND | Top-2 | **0.9595** | **0.9793 (0.8951,1.0)** | **0.9735** | **1.3575** | **1.4091** | **2.5563** | **36.6339** |
| | | Top-3 | 0.9490 | 0.9738 (0.8794,1.0) | 0.9625 | 1.5307 | 1.6137 | 3.0576 | 38.1265 |
| | Bitwise-OR | Top-2 | 0.9490 | 0.9738 (0.8794,1.0) | 0.9625 | 1.5307 | 1.6137 | 3.0576 | 38.1265 |
| | | Top-3 | 0.9490 | 0.9738 (0.8794,1.0) | 0.9625 | 1.5307 | 1.6137 | 3.0576 | 38.1265 |

The model trained on the P CXR dataset was considered the baseline. From Table 6, for the P1 test group, we observed that the performance achieved through bitwise operations applied to the predictions of the top-K adult models did not outperform the baseline. The baseline performance was significantly superior to other adult models ($p < 0.05$). For the P2 test group, we observed that the performance achieved through Bitwise-AND applied to the predictions of the top-2 adult models outperformed the baseline in all evaluation metrics. A similar performance improvement was observed for the test group P3 using a bitwise-AND applied to the predictions of the top-2 adult models. However, no statistical significance was observed ($p > 0.05$) in the reported Dice scores for the P2 and P3 test groups.

We constructed a stacked ensemble of the top-K (K = 2, 3) adult models as discussed in Section 3.3. The performance achieved through the stacked ensemble is shown in Table 7 and was compared to the baseline. Fig. 7 shows the predictions of the stacked ensemble for CXR instances from the P1, P2, and P3 test groups.

Table 7 – Performance achieved through a stacked ensemble of the top-K (K = 2, 3) adult models for each pediatric test group. Bold numerical values denote superior performance in each test group.

| Test groups | Models | IoU | Dice (CI) | MS-SSIM | MLCD | AHS | HD95 | ASSD |
|---|---|---|---|---|---|---|---|---|
| P1 | P | 0.8637 | 0.9268 (0.7728,1.0) | 0.9745 | 3.4513 | 7.0228 | 39.3002 | 41.2564 |
| | Top-2 | **0.8823** | **0.9375 (0.7944,1.0)** | **0.9774** | **3.2382** | **3.3637** | **36.0285** | **40.1534** |
| | Top-3 | 0.8774 | 0.9347 (0.7887,1.0) | 0.9761 | 5.4627 | 3.9091 | 37.5983 | 41.0432 |
| P2 | P | 0.9004 | 0.9476 (0.8416,1.0) | 0.9360 | 4.5234 | 4.8677 | 29.1362 | 52.9087 |
| | Top-2 | **0.9178** | **0.9571 (0.8607,1.0)** | **0.9565** | **2.1422** | **2.7942** | **20.1013** | **44.2035** |
| | Top-3 | 0.9161 | 0.9562 (0.8589,1.0) | 0.9556 | 2.4771 | 3.2942 | 21.3986 | 45.0335 |
| P3 | P | 0.9330 | 0.9654 (0.8573,1.0) | 0.9390 | 4.1418 | 2.4546 | 16.8817 | 60.2842 |
| | Top-2 | 0.9345 | 0.9661 (0.8591,1.0) | 0.9414 | 5.6276 | 2.7500 | 3.8712 | 39.6186 |
| | Top-3 | **0.9409** | **0.9695 (0.8678,1.0)** | **0.9436** | **3.2434** | **2.2046** | **2.3517** | **34.6828** |

As observed from Table 7 and Fig. 7, the stacked ensemble of the top-2 adult models outperformed the baseline for the pediatric test group P1. This was also evident from the MS-SSIM quality maps (Fig. 8(a) bottom) that showcased increased red activations at varying scales for the stacked ensemble predictions compared to the baseline (Fig. 8(a) top). This performance improvement could be attributed to the CXR modality-specific weight initializations as discussed in the literature (Islam et al., 2017; Rajaraman et al., 2018, 2020; Zamzmi et al., 2022; Zamzmi, Rajaraman, & Antani, 2021). Modality-specific weight initializations might have facilitated transferring CXR modality-specific knowledge that was fine-tuned to improve performance in the related pediatric lung segmentation task. A similar improvement in performance was observed for the P2 and P3 test groups. For the P2 test group, the stacked ensemble of the top-2 adult models outperformed the baseline in all evaluation metrics. For the P3 test group, the stacked ensemble of the top-3 adult models delivered superior performance in terms of all evaluation metrics. The MS-SSIM quality maps ((Fig. 8(b) bottom) and (Fig. 8(c) bottom)) showcased qualitative shreds of evidence of this performance improvement with increasing red activations at varying scales for the masks predicted by the stacked ensemble compared to the baseline (Fig. 8(b) top and Fig. 8(c) top, respectively).

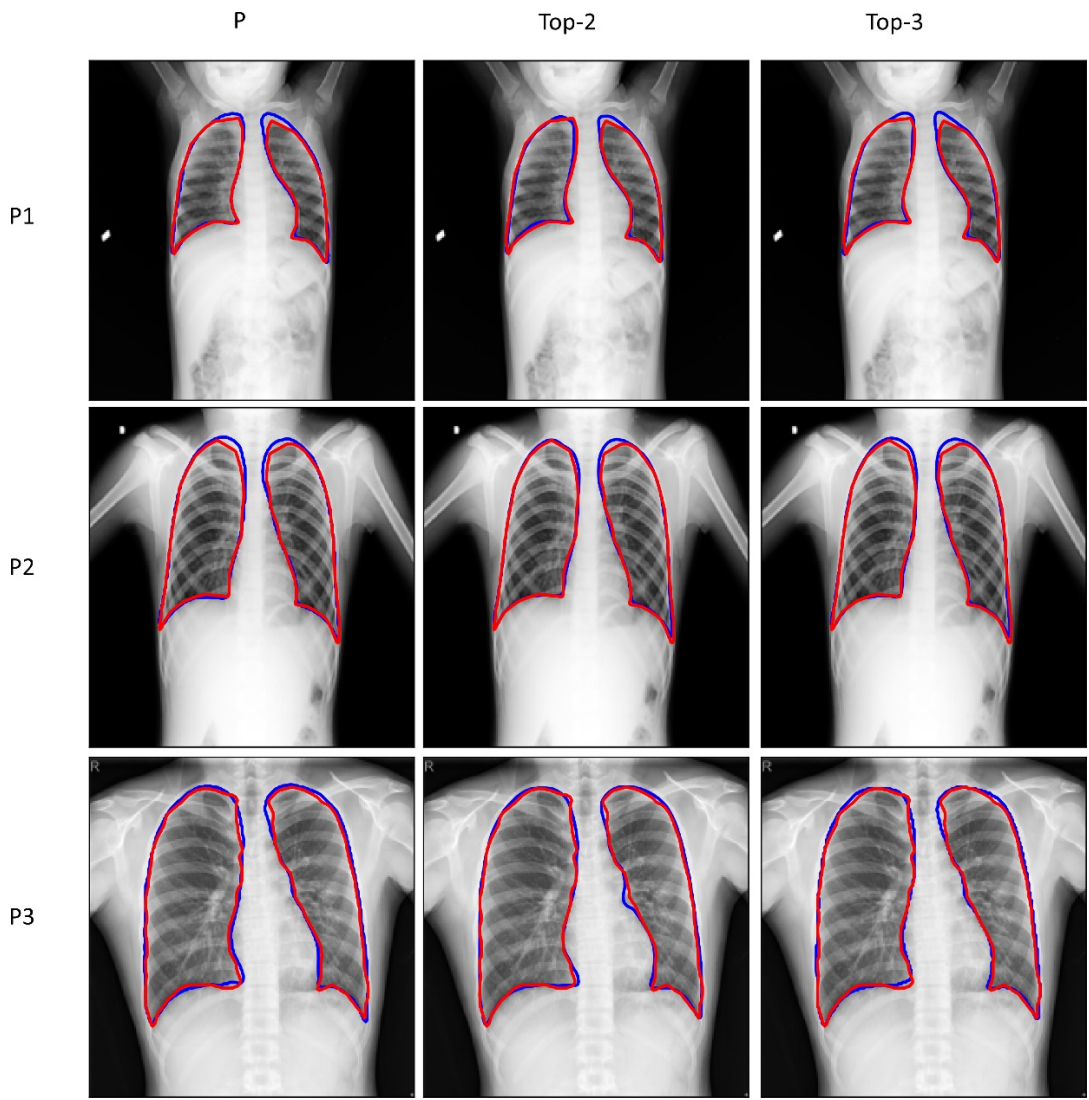

Fig. 7 – Visualizing and comparing the predictions of the Inception-V3 model trained on the P CXR dataset (baseline), the stacked ensemble of the top-2 and top-3 adult models, respectively, for a CXR instance from P1 (age = 18 months), P2 (age = 10

years), and P3 (age = 15 years) test groups. The red and blue contours denote GT and predictions, respectively.

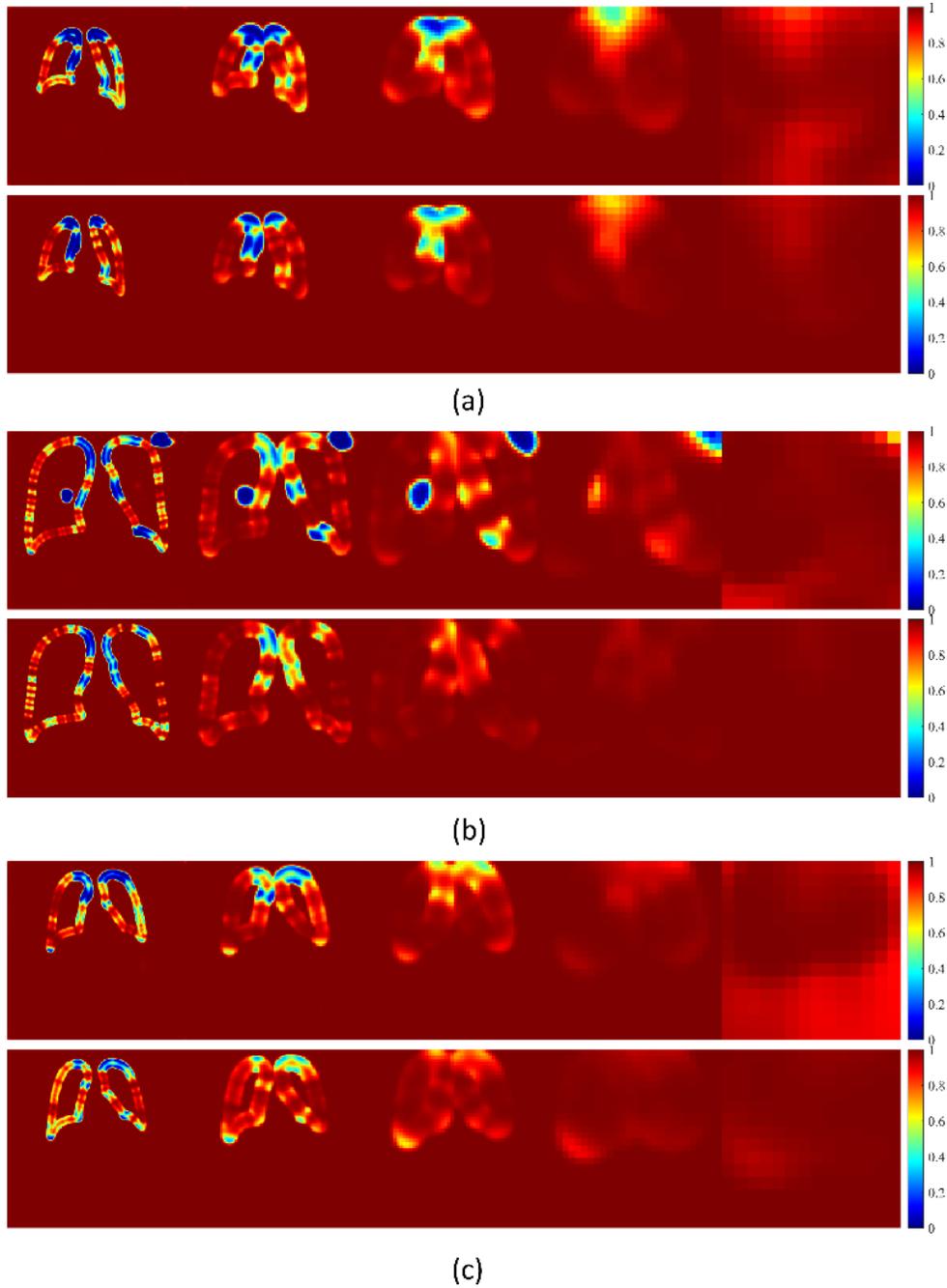

Fig. 8 – MS-SSIM quality maps at five different scales. (a) The predictions of the Inception-V3-UNet model trained on the P CXR dataset (top) and the stacked ensemble of the top-2 adult models (bottom) for an instance of CXR (age = 10 months) from the test group P1. (b) The prediction of the Inception-V3-UNet model trained on the P CXR dataset (top) and the stacked ensemble of the top-2 adult models (bottom) for a CXR instance (age = 3 years) from test group P2. (c) The prediction of the Inception-V3-UNet model trained on the P CXR dataset (top) and the stacked ensemble of the top-3 adult models (bottom) for an instance of CXR (age = 15 years) from test group P3.

We further constructed an ensemble of stacked ensembles as follows: The predictions of the stacked ensemble of the top-2 and top-3 adult models, respectively, were aggregated using bitwise operations. Table 8 shows the performance achieved in this regard.

Table 8 – Performance achieved through an ensemble of stacked ensembles. Bold numerical values denote superior performance in each test group. The * denotes statistical significance ($p < 0.05$).

| Test groups | Method | IoU | Dice (CI) | MS-SSIM | MLCD | AHS | HD95 | ASSD |
|---|---|---|---|---|---|---|---|---|
| P1 | P | 0.8637 | 0.9268 (0.7728,1.0) | 0.9745 | 3.4513 | 7.0228 | 39.3002 | 41.2564 |
| | Bitwise-AND | **0.8862** | **0.9397 (0.799,1.0)** | **0.9809** | **3.2002** | **2.7273** | **35.1212** | **38.9874** |
| | Bitwise-OR | 0.8702 | 0.9306 (0.7804,1.0) | 0.9709 | 6.2282 | 4.1819 | 38.1584 | 40.9191 |
| P2 | P | 0.9004 | 0.9476 (0.8416,1.0) | 0.9360 | 4.5234 | 4.8677 | 29.1362 | 52.9087 |
| | Bitwise-AND | **0.9231** | **0.9600 (0.8668,1.0)*** | **0.9607** | **2.0452** | **2.6618** | **19.2138** | **43.1025** |
| | Bitwise-OR | 0.9106 | 0.9532 (0.8527,1.0) | 0.9490 | 2.4146 | 3.1324 | 22.8723 | 46.1344 |
| P3 | P | 0.9330 | 0.9654 (0.8573,1.0) | 0.9390 | 4.1418 | 2.4546 | 16.8817 | 60.2842 |
| | Bitwise-AND | 0.9189 | 0.9577 (0.8387,1.0) | 0.9357 | 5.8449 | 2.5910 | 20.4863 | 43.1033 |
| | Bitwise-OR | **0.9359** | **0.9669 (0.8611,1.0)** | **0.9467** | **3.0849** | **2.0728** | **3.6212** | **39.1123** |

For the P1 test group, we observed that the performance achieved with the bitwise-AND of the predictions of the stacked ensemble of the top-2 and top-3 adult models is superior considering all evaluation metrics compared to the baseline and those achieved by the top-2 and top-3 stacked ensemble models individually. A similar improvement in performance compared to the baseline and the individual top-2 and top-3 stacked ensemble models were observed for the P2 test group through the bitwise-AND operation. For the P3 test group, we observed the bitwise-OR delivered superior performance compared to the baseline. However, this performance improvement was not significant ($p > 0.05$) compared to that achieved by the stacked ensemble of the top-3 adult models (Table 7).

## 5. CONCLUSION AND SCOPE FOR FUTURE RESEARCH

The strength of this study lies in proposing an effective framework to improve the generalization of adult lung segmentation models to the unseen pediatric population across the various age groups. This is achieved using a combination of approaches consisting of CXR modality-specific weight initializations, a stacked ensemble, and an ensemble of stacked ensembles. Our proposal does not require any additional pediatric CXRs data or associated labels to improve cross-domain generalization, allowing the expansion and utilization of existing systems for pediatric screening.

Although our results are promising, this study has some limitations. First, the limited number of publicly available adult CXRs, the pediatric CXRs, and their associated lung masks. Our choice of models was limited to the widely used UNet and TransUNet models. Advanced architecture including LinkNet (Chaurasia & Culurciello, 2018), PSPNet (H. Zhao et al., 2017), and Tri-lateral Attention Net (Zamzmi, Rajaraman, Sachdev, et al., 2021), among others, and their ensembles might reveal other insights. Future research could focus on continuous domain generalization, i.e., generalizing to a stream of external non-stationary targets. Interpretable domain generalization methods could also be proposed to explain the domain-variant and domain-invariant features.

**FUNDING**

This work is supported by the Intramural Research Program (IRP) of the National Library of Medicine (NLM), and the National Institutes of Health (NIH).

**CONFLICTS OF INTEREST**

The authors declare no conflict of interest.

**AUTHOR CONTRIBUTIONS**

Sivaramakrishnan Rajaraman: Conceptualization, Methodology, Data curation, Software, Visualization, Formal analysis, Writing- Original draft preparation, Writing- Reviewing and Editing. Feng Yang: Conceptualization, Data curation, Methodology, Software, Writing- Reviewing and Editing; Ghada

Zamzmi: Conceptualization, Methodology, Writing- Reviewing and Editing. Zhiyun Xue:

Conceptualization, Methodology, Writing- Reviewing and Editing. Sameer Antani: Conceptualization,

Methodology, Investigation, Resources, Supervision, Funding acquisition, Project administration, Writing-

Reviewing, and Editing.

# REFERENCES

Altman, D. G., & Bland, J. M. (2011). Statistics notes: How to obtain the P value from a confidence interval. *BMJ (Online)*, *343*(7825), 1–2. https://doi.org/10.1136/bmj.d2304

Arthur, R. (2000). Interpretation of the paediatric chest X-ray. *Paediatric Respiratory Reviews*, *1*(1), 41–50. https://doi.org/10.1053/prrv.2000.0018

Aydin, O. U., Taha, A. A., Hilbert, A., Khalil, A. A., Galinovic, I., Fiebach, J. B., Frey, D., & Madai, V. I. (2021). An evaluation of performance measures for arterial brain vessel segmentation. *BMC Medical Imaging*, *21*(1), 1–12. https://doi.org/10.1186/s12880-021-00644-x

Candemir, S., & Antani, S. (2019). A review on lung boundary detection in chest X-rays. In *International Journal of Computer Assisted Radiology and Surgery*. https://doi.org/10.1007/s11548-019-01917-1

Candemir, S., Antani, S., Jaeger, S., Browning, R., & Thoma, G. (2015). Lung Boundary Detection in Pediatric Chest X-rays. *Medical Imaging 2015: Pacs and Imaging Informatics: Next Generation and Innovations*. https://doi.org/10.1117/12.2081060

Candemir, S., Jaeger, S., Antani, S., Bagci, U., Folio, L. R., Xu, Z., & Thoma, G. (2016). Atlas-based rib-bone detection in chest X-rays. *Computerized Medical Imaging and Graphics*. https://doi.org/10.1016/j.compmedimag.2016.04.002

Chaurasia, A., & Culurciello, E. (2018). LinkNet: Exploiting encoder representations for efficient semantic segmentation. *2017 IEEE Visual Communications and Image Processing, VCIP 2017*, *2018-Janua*, 1–4. https://doi.org/10.1109/VCIP.2017.8305148

Chen, J., Lu, Y., Yu, Q., Luo, X., Adeli, E., Wang, Y., Lu, L., Yuille, A. L., & Zhou, Y. (2021). TransUNet: Transformers Make Strong Encoders for Medical Image Segmentation. *ArXiv Preprint*, 1–13. http://arxiv.org/abs/2102.04306

Cohen, J. P., Hashir, M., Brooks, R., & Bertrand, H. (2020). On the limits of cross-domain generalization in automated X-ray prediction. *Proceedings of Machine Learning Research*, *121*, 136–149. http://arxiv.org/abs/2002.02497

Hesamian, M. H., Jia, W., He, X., & Kennedy, P. (2019). Deep Learning Techniques for Medical Image Segmentation: Achievements and Challenges. *Journal of Digital Imaging*, *32*(4), 582–596. https://doi.org/10.1007/s10278-019-00227-x

Iakubovskii, P. (2019). *Segmentation Models*. https://github.com/qubvel/segmentation_models; GitHub.

Islam, M. T., Aowal, M. A., Minhaz, A. T., & Ashraf, K. (2017). Abnormality Detection and Localization in Chest X-Rays using Deep Convolutional Neural Networks. *ArXiv*. http://arxiv.org/abs/1705.09850

Jaeger, S., Candemir, S., Antani, S., Wang, Y.-X. J., Lu, P.-X., & Thoma, G. (2014). Two public chest X-ray datasets for computer-aided screening of pulmonary diseases. *Quantitative Imaging in Medicine and Surgery*, *4*(6), 475–477. https://doi.org/10.3978/j.issn.2223-4292.2014.11.20

Lyu, J., Zhang, Y., Huang, Y., Lin, L., & Cheng, P. (2022). AADG : Automatic Augmentation for Domain Generalization on Retinal Image Segmentation. *IEEE Transactions on Medical Imaging*, *41*(12), 3699–3711.

Oda, S., Awai, K., Suzuki, K., Yanaga, Y., Funama, Y., MacMahon, H., & Yamashita, Y. (2009). Performance of radiologists in detection of small pulmonary nodules on chest radiographs: Effect of rib suppression with a massive-training artificial neural network. *American Journal of Roentgenology*. https://doi.org/10.2214/AJR.09.2431

Osadebey, M., Andersen, H. K., Waaler, D., Fossaa, K., Martinsen, A. C. T., & Pedersen, M. (2021). Three-stage segmentation of lung region from CT images using deep neural networks. *BMC Medical Imaging*, *21*(1), 1–19. https://doi.org/10.1186/s12880-021-00640-1

Ou, Y., & Rhee, K. H. (2010). A survey on image hashing for image authentication. *IEICE Transactions on Information and Systems*, *E93-D*(5), 1020–1030. https://doi.org/10.1587/transinf.E93.D.1020

Rajaraman, S., Candemir, S., Kim, I., Thoma, G., Antani, S., Rajaraman, S., Candemir, S., Kim, I., Thoma, G., & Antani, S. (2018). Visualization and Interpretation of Convolutional Neural Network Predictions in Detecting Pneumonia in Pediatric Chest Radiographs. *Applied Sciences*, *8*(10), 1715. https://doi.org/10.3390/app8101715

Rajaraman, S., Kim, I., & Antani, S. K. (2020). Detection and visualization of abnormality in chest radiographs using modality-specific convolutional neural network ensembles. *PeerJ*. https://doi.org/10.7717/peerj.8693

Rajaraman, S., Yang, F., Zamzmi, G., Xue, Z., & Antani, S. K. (2022). A Systematic Evaluation of Ensemble Learning Methods for Fine-Grained Semantic Segmentation of Tuberculosis-Consistent Lesions in Chest Radiographs. *Bioengineering*, *9*(9), 413. https://doi.org/10.3390/bioengineering9090413

Ronneberger, O., Fischer, P., & Brox, T. (2015). U-net: Convolutional networks for biomedical image segmentation. *Lecture Notes in Computer Science (Including Subseries Lecture Notes in Artificial Intelligence and Lecture Notes in Bioinformatics)*. https://doi.org/10.1007/978-3-319-24574-4_28

Stacke, K., Eilertsen, G., Unger, J., & Lundström, C. (2021). Measuring Domain Shift for Deep Learning in Histopathology. *IEEE Journal of Biomedical and Health Informatics*, *25*(2), 325–336.

Stirenko, S., Kochura, Y., Alienin, O., Rokovyi, O., Gordienko, Y., Gang, P., & Zeng, W. (2018). Chest X-Ray Analysis of Tuberculosis by Deep Learning with Segmentation and Augmentation. *2018 IEEE 38th International Conference on Electronics and Nanotechnology, ELNANO 2018 - Proceedings*. https://doi.org/10.1109/ELNANO.2018.8477564

Suzuki, K. (2017). Overview of deep learning in medical imaging. In *Radiological Physics and Technology* (Vol. 10, Issue 3, pp. 257–273). https://doi.org/10.1007/s12194-017-0406-5

Therrien, R., & Doyle, S. (2018). *Role of training data variability on classifier performance and generalizability*. *1058109*(March 2018), 5. https://doi.org/10.1117/12.2293919

Wolpert, D. H. (1992). Stacked generalization. *Neural Networks*, *5*(2), 241–259. https://doi.org/10.1016/S0893-6080(05)80023-1

Xin, K. Z., Li, D., & Yi, P. H. (2022). Limited generalizability of deep learning algorithm for pediatric pneumonia classification on external data. *Emergency Radiology*, *29*(1), 107–113. https://doi.org/10.1007/s10140-021-01954-x

Yang, F., Lu, P. X., Deng, M., Xi, Y., W, J., Rajaraman, S., Xue, Z., Folio, L. R., Antani, S. K., & Jaeger, S. (2022). Annotations of Lung Abnormalities in the Shenzhen Chest Pulmonary Diseases. *MDPI Data*, 1–5.

Yeung, M., Yang, G., Sala, E., Schönlieb, C.-B., & Rundo, L. (2021). *Incorporating Boundary Uncertainty into loss functions for biomedical image segmentation*. http://arxiv.org/abs/2111.00533

Zamzmi, G., Rajaraman, S., & Antani, S. (2021). UMS-Rep: Unified modality-specific representation for efficient medical image analysis. *Informatics in Medicine Unlocked*, *24*(March), 100571. https://doi.org/10.1016/j.imu.2021.100571

Zamzmi, G., Rajaraman, S., Hsu, L.-Y., Sachdev, V., & Antani, S. (2022). Real-time echocardiography image analysis and quantification of cardiac indices. *Medical Image Analysis*, *80*, 102438. https://doi.org/10.1016/j.media.2022.102438

Zamzmi, G., Rajaraman, S., Sachdev, V., & Antani, S. (2021). Trilateral Attention Network for Real-Time Cardiac Region Segmentation. *IEEE Access*, *9*, 118205–118214. https://doi.org/10.1109/ACCESS.2021.3107303

Zech, J. R., Badgeley, M. A., Liu, M., Costa, A. B., Titano, J. J., & Oermann, E. K. (2018). Variable generalization performance of a deep learning model to detect pneumonia in chest radiographs: A cross-sectional study. *PLoS Medicine*, *15*(11), 1–17. https://doi.org/10.1371/journal.pmed.1002683

Zhang, R., Xu, Q., Huang, C., Zhang, Y., & Wang, Y. (2022). Semi-Supervised Domain Generalization for Medical Image Analysis. *Proceedings - International Symposium on Biomedical Imaging*, *2022-March*. https://doi.org/10.1109/ISBI52829.2022.9761561

Zhao, H., Shi, J., Qi, X., Wang, X., & Jia, J. (2017). Pyramid scene parsing network. *Proceedings - 30th IEEE Conference on Computer Vision and Pattern Recognition, CVPR 2017*, *2017-Janua*, 6230–6239. https://doi.org/10.1109/CVPR.2017.660

Zhao, S., Wu, B., Chu, W., Hu, Y., & Cai, D. (2019). *Correlation Maximized Structural Similarity Loss for Semantic Segmentation*. http://arxiv.org/abs/1910.08711

Zheng, J., Lu, C., Hao, C., & Chen, D. (2021). Improving the Generalization Ability of Deep Neural Networks for Cross-Domain

Visual Recognition. *IEEE Transactions on Cognitive and Developmental Systems*, *13*(3), 607–620.